\documentclass[
    reprint,
    showpacs,
    preprintnumbers,
    amsfonts,
    amssymb,
    amsmath,
    aps,
    prd
    ]{revtex4-1}

\usepackage[latin1]{inputenc}
\usepackage[T1]{fontenc}
\usepackage[english]{babel}
\usepackage[pdftex]{xcolor}
\usepackage{graphicx}
\usepackage{xspace}
\usepackage{tabularx}

\definecolor{darkred}{rgb}{0.5,0,0}
\definecolor{darkgreen}{rgb}{0,.5,0}
\definecolor{darkblue}{rgb}{0,0,.5}
\usepackage[pdftex,bookmarksopen]{hyperref}
\hypersetup{
  pdftitle={An explicit SU(12) family and flavor unification model with natural
    fermion masses and mixings}, 
  pdfauthor={Robert Feger}, 
  pdfsubject={}, 
  pdfcreator={pdfTeX},
  pdfproducer={Robert Feger}, 
  pdfkeywords={},
  colorlinks=true,        
  linkcolor=darkblue,      
  citecolor=darkblue,    
  urlcolor=darkblue,      
  pdfborder={0 0 0}
}

\renewcommand{\epsilon}{\varepsilon}
\DeclareMathOperator{\diag}{diag}
\newcommand\order[1]{\mathcal{O}(#1)}
\newcommand\abs[1]{\lvert#1\rvert}
\newcommand{\SU}[1]{\ensuremath{{\text{SU}(#1)}}}
\newcommand{\U}[1]{\ensuremath{{\text{U}(#1)}}}

\newcommand{\vev}[1]{\langle #1 \rangle}

\newcommand\MeV{\,\text{MeV}}
\newcommand\meV{\,\text{meV}}
\newcommand\GeV{\,\text{GeV}}
\newcommand\eVsq{\,\text{eV}^2}

\newcommand{\irrep}[1]{\ensuremath{\boldirrep{#1}}}

\newcommand{\boldirrep}{\mathbf}
\newlength{\irrepwidth}
\newlength{\irrepbarthickness}
\newlength{\irrepbarheight}
\newcommand{\irrepbar}[1]{%
	\settoheight{\irrepbarheight}{\ensuremath{\boldirrep{#1}}}%
	\settowidth{\irrepwidth}{\ensuremath{\boldirrep{#1}}}%
	\makebox[0pt][l]{\ensuremath{\boldirrep{#1}}}%
	\rule[1.2\irrepbarheight]{\irrepwidth}{\irrepbarthickness}%
}
\newcommand{\irrepvar}[1]{(\irrep{#1})}
\newcommand{\irrepbarvar}[1]{(\irrepbar{#1})}

\newcommand{\irrepsub}[2]{\ensuremath{\irrep{#1}_\text{#2}}}
\newcommand{\irrepbarsub}[2]{\ensuremath{\irrepbar{#1}_\text{#2}}}

\newcommand\fermion\irrepsub
\newcommand\fermionbar\irrepbarsub

\newcommand{\higgs}[1]{\irrepsub{#1}{H}}
\newcommand{\higgsbar}[1]{\irrepbarsub{#1}{H}}

\newcommand\twocolumnsbold[1]{\multicolumn{2}{l}{\textbf{#1}}}
\newcommand{\massivefermionpair}[2]{%
\ensuremath{#1{\times}#2}%
}
\newcommand{\UpType}[2]{\text{\textbf{U#1#2}:}}
\newcommand{\DownType}[2]{\text{\textbf{D#1#2}:}}
\newcommand{\Majorana}[2]{\text{\textbf{MN#1#2}:}}
\newcommand{\Dirac}[2]{\text{\textbf{DN#1#2}:}}
\newcommand{\ldot}{{.}}

\newcommand\hu[1]{h^\text{u}_{#1}}
\newcommand\hd[1]{h^\text{d}_{#1}}
\newcommand\hl[1]{h^\ell_{#1}}
\newcommand\hmn[1]{{h^\text{mn}_{#1}}}
\newcommand\hdn[1]{{h^\text{dn}_{#1}}}

\newcommand\modelgroup{\SU{12}\xspace}

\begin{document}

\preprint{FERMILAB-PUB-12-105-T}

\title{\boldmath%
    An explicit \modelgroup family and flavor unification model with\\
    natural fermion masses and mixings%
}

\date{April 24, 2012}

\author{Carl H. Albright}
\email{albright@fnal.gov}
\affiliation{
    \mbox{Department of Physics, Northern Illinois University, DeKalb, IL 60115}\\
    and\\
    Theoretical Physics, Fermilab, Batavia, IL 60510
}

\author{Robert P. Feger}
\email{robert.feger@vanderbilt.edu}
\author{Thomas W. Kephart}
\email{thomas.w.kephart@vanderbilt.edu}
\affiliation{Department of Physics and Astronomy, Vanderbilt University,
  Nashville, TN 37235}

\begin{abstract}
We present an \modelgroup unification model with three light chiral families,
avoiding any external flavor symmetries. The hierarchy of quark and lepton
masses and mixings is explained by higher dimensional Yukawa interactions
involving Higgs bosons that contain \SU{5} singlet fields with VEVs about 50 times
smaller than the \modelgroup unification scale. The presented model has been
analyzed in detail and found to be in very good agreement with the observed
quark and lepton masses and mixings.
\end{abstract}


\maketitle
\enlargethispage{-12pt}
\section{Introduction}
\label{sec:Introduction}
The elementary fermions in the Standard Model (SM) appear in three families,
which are a triplication according to their gauge transformations in the
unbroken electroweak Lagrangian. The masses vary by several orders of magnitude
and are a mystery within the SM. A wide range of models introduce a spontaneously
broken flavor symmetry, with the associated group being either continuous or
discrete~\cite{Altarelli:2010gt, Ishimori:2010au}. Different charges assigned
to the families account for the mass and mixing hierarchy by producing
different mass terms with an appropriate Higgs sector. This is especially
necessary in SO(10) Grand Unified Theories (GUTs), where the \irrep{16} spinor
irreducible representation (irrep) is the only complex representation
yielding chiral fermions but no exotic fermions~\cite{Albright:1998vf}. Early on,
unification groups based on higher rank orthogonal groups such as SO(18) were
explored~\cite{GellMann:1980vs,Fujimoto:1981bv}, but the number of exotic
fields introduced became prohibitive. Early studies of the case of \SU{N}
family symmetry include models based on \SU{11}~\cite{Georgi:1979md,Kim:1981bb}
and \SU{9}~\cite{Frampton:1979cw,Frampton:1979fd,Frampton:2009ce}.

Grand Unified Theories based on the groups \SU{N} with $N{>}5$ can give rise to
a different approach: while all families transform in the same way under the SM
gauge group it is possible to assign them to different antisymmetric multiplets
of \SU{N} to obtain a non-trivial flavor structure. Since in \SU{5} a family
can only be assigned to \irrep{10}+\irrepbar{5} (or to the conjugated pair)
\cite{Georgi:1974sy}, the unification group must be larger, hence $N{>}5$. This
idea has led to the supersymmetric \SU{7}~\cite{Barr:2008gz} and the
non-supersymmetric \SU{8} mo\-dels~\cite{Barr:2008pn} proposed by Barr. In a
previous publication~\cite{ Dent:2009pd} two of the present authors (RF and TWK) and
others have constructed a hybrid of the latter two approaches, with a partial
assignment to different irreps of \SU{9} and four discrete symmetries in a
non-supersymmetric model. Since then we have developed a systematic scan of
\SU{N}'s that loops over all possible fermion assignments to find viable models
with or without discrete symmetries, including the hybrid case mentioned above.
We present here an \modelgroup model found by this scan which is free of any
imposed external flavor symmetries. We have now also included the assignment of
right-handed neutrinos, which allows the analysis of the full lepton sector as
well, which is more ambitious than~\cite{Barr:2008gz, Barr:2008pn,Dent:2009pd}.

In Sec.~\ref{sec:ModelConstructionPrincipal} we outline the
construction of the model by effective higher dimensional operators, which
produce the mass and mixing hierarchy. Sec.~\ref{sec:ModelScan} gives a
brief survey of the model scan procedure which enabled us to find the
\SU{12} model presented here. Sec.~\ref{sec:ModelPresentation}
is the major section and presents the \modelgroup model in detail: In
Sec.~\ref{ssec:ThreeFamiliesEmbedding} we demonstrate how three chiral
families arise from \modelgroup in our model. After listing the fermion
assignments and the Higgs sector in Sec.~\ref{ssec:Assignments} we
construct the Yukawa interactions for the quark and lepton sector in
Sec.~\ref{ssec:YukawaInteractions}, compute the resulting mass matrices,
which involves the seesaw mechanism for the neutrinos, and finally fit the
mass matrices to the measured values of the known masses and mixings in
Sec.~\ref{ssec:Phenomenology}. The discussion of the results
and implications are presented in Sec.~\ref{sec:Discussion}. We summarize
and conclude in Sec.~\ref{sec:SummaryAndConclusion}.\newpage

\section{Fermion Mass Hierarchy from Higher Dimensional Operators}
\label{sec:ModelConstructionPrincipal}

The Yukawa couplings in the Standard Model correctly para\-metrize the observed
masses and mixings of quarks and leptons, yet the SM fails to explain why the
coupling strengths are spread over a range of five orders of magnitude. Assuming
an underlying naturalness of the Yukawa couplings, one can understand their
measured values in an effective field theory scenario as coefficients of
effective operators encoding short distance physics above a scale $\Lambda$. In
these effective theories only heavy fermions obtain their masses from
renormalizable, four-dimensional Yukawa couplings, while the masses of the
lighter fermions are due to higher dimensional operators. In
non-super\-symmetric models these operators may stem from loops involving Higgs
fields, while in supersymmetric models these loops are suppressed by factors of
$M_\text{SUSY}/M_\text{GUT}$. In the latter case the masses of the lighter
fermions must come from tree-level diagrams at the $M_\text{GUT}$ scale, which
we have pursued in the construction of the model presented here. However, we do
not consider the phenomenology of the supersymmetric partners of Standard Model
particles by simply assuming that supersymmetry is broken at a scale high enough
to be inaccessible to current collider experiments, but low enough not to upset
the suppression of loops. Here we expect supersymmetry breaking in the
$10^8$--$10^{10}\GeV$ range, which would soften but not solve the hierarchy
problem.

To this end we introduce vectorlike heavy fermions with masses $M$ at the
\modelgroup unification scale and extend the Higgs sector by introducing
\modelgroup Higgs bosons containing \SU{5} singlet vacuum expectation
values (VEVs). These allow one to construct Froggatt-Nielsen-type diagrams
\cite{Froggatt:1978nt}, i.\,e.\ tree-level diagrams with heavy fermions as
one or more mass insertions and Higgs bosons containing \SU{5} singlet VEVs
(see e.\,g.~\cite{Hall:1985dx}), which are assumed to be about 50 times
lighter than the \modelgroup unification scale. In going to the electroweak
scale or lower, these mass insertions can be integrated out leaving effective
Yukawa couplings involving Higgs bosons with electroweak VEVs and \SU{5} singlet
VEVs, suppressed by the masses of the heavy fermions at the \modelgroup
unification scale, $M_{\modelgroup}$. After breaking the Higgs sector to
\SU{5} and subsequently to $G_\text{SM}$, the \SU{5} singlet VEVs
$\vev{1}_\SU{5}$ and \modelgroup unification scale $M_{\modelgroup}$ appear
in the ratio:
\begin{equation}\label{eq:DefinitionEpsilon}
	\epsilon = \frac{\vev{1}_\SU{5}}{M_{\modelgroup}}\sim \frac{1}{50}.
\end{equation}
Yukawa interactions of dimension $4+n$ give rise to mass matrix elements of
the form:
\begin{equation}
	h_{ij}\epsilon^n v\, u^T_{iL}u^c_{jL},
\end{equation}
where $h_{ij}$ are the Yukawa couplings and $v{=}174\GeV$ is the electroweak VEV.
The dimensionless quantity $\epsilon$ parametrizes the mass and mixing
hierarchy in our model. The power $n$ of $\epsilon$ in each Yukawa interaction
is the number of mass insertions and \SU{5} singlet VEVs and represents an
order of the effective operator higher than four. The value is roughly the
ratio of the bottom-quark mass to the top-quark mass. The assumption in our
model is that all other mass and mixing ratios can be expressed in powers of
$\epsilon$, while the Yukawa couplings $h_{ij}$ are of $\order{1}$ at the
\modelgroup unification scale, with the dimension of the corresponding
effective Yukawa interaction chosen accordingly.

The top-quark Yukawa coupling in the Standard Model is of order unity,
suggesting that the renormalizable, dimension four interaction is the correct
description. Since all other quark masses are small compared to the top-quark
mass they must arise from higher dimensional Yukawa couplings. The up-type
quarks exhibit an especially strong mass hierarchy compared to the down-type
quarks. The mixing angles of the CKM matrix are small, leading to similar up-
and down-type mass matrices, but with somewhat stronger hierarchies for the
former. The neutrinos on the other hand have comparable masses and large mixing
angles, leading to a light neutrino mass matrix with either a mild or little
hierarchy, while the charged leptons exhibit a strong mass hierarchy.

\section{Model Search}
\label{sec:ModelScan}

The \modelgroup model presented in this paper was found by a computer program
developed by one of us (RPF) to scan models of the type described in
Sec.~\ref{sec:ModelConstructionPrincipal}. The scan essentially seeks models by
brute force, i.\,e.,\ constructing all possible combinations of fermion
assignments, Higgs irreps, and massive fermions and probing them for their
phenomenological implications.

Due to the enormous number of combinations, the scan is constructed in five
enclosing loops for a specific \SU{N} group being searched: The first loop runs
over anomaly-free sets of irreps that yield three chiral families at the \SU{5}
level. The fermions embedded in \irrep{10}'s of \SU{5} are assigned to these
sets of irreps first, which includes all up-type quark fields. This is
sufficient to compute the up-type mass matrix, once the Higgs irreps and massive
fermions are defined. All subsets of three of the anomaly-free, three-family
sets of \SU{N} irreps can be assigned to fermions of \SU{5}\ \irrep{10}'s, which
constitutes the second loop.

For each of these assignments, a third loop over all combinations of Higgs
irreps and massive fermions taken from a basic set, is performed that
computes the orders of the up-type mass matrix elements for each combination.
Imposed requirements for the ordering can already filter out bad combinations
of fermion assignments, Higgs and massive fermions. A crucial requirement is
that only the top-quark mass term is of dimension 4, i.\,e.\,,\ of zeroth
order in $\epsilon$.

For each of these filtered combinations, there is an analogous fourth loop over
all assignments of fields embedded in the \SU{5}\ \irrepbar{5}'s to irreps of
the anomaly-free, three-family set the outer loop is currently investigating.
Since assignments for the \irrep{10}'s and \irrepbar{5}'s for all three
families and definitions of Higgs irreps as well as massive fermions is
sufficient to compute the orders of the up-type and down-type mass matrix
elements and thus the CKM matrix, a fit of the prefactors of the up and down
quark mass matrices and thus the complete quark sector is possible.

Having singled out quark models with reasonable phenomenology, a fifth loop
adds assignments of right-handed neutrinos. For each of these assignments the
orders of Dirac- and Majorana-neutrino mass matrix elements are computed,
as well as the corresponding light-neutrino mass matrix via the type I seesaw
mechanism. Together with the charged-lepton mass matrix, which is the
transpose of the down-type mass matrix, the neutrino masses and mixings can
be calculated.

For a fit to neutrino data analogous to the quark sector, only the mass
differences squared are used, which allows for either a normal or an inverted
hierarchy. The overall fit performed at this stage is a combined fit including
the quark mass matrices as well. The fit yields a $\chi^2$ value for each
model, which allows one to select suitable models automatically. A second
requirement imposed is that the prefactors be of order one. A more precise
description of these steps and their results will be published in a follow-up
paper by the authors.

\section{Model Properties}
\label{sec:ModelPresentation}

As a result of both the computer scan and by comparing many models by hand we
have found a set of models of considerable interest.
Here we will choose one specific \SU{12} example to explore, which has
many attractive features. Out of the thousands of models we have studied there
is a large handful that fit the data quite well. Hence, our \SU{12} model is
neither generic nor unique.

\subsection{Three Families in SU(12)}
\label{ssec:ThreeFamiliesEmbedding}

As a prime example of our procedure, we begin with the set of \modelgroup
irreps
\begin{equation}\label{eq:AnomFreeSet}
    6\irrepvar{495} + 4\irrepbarvar{792} + 4\irrepbarvar{220} +
    \irrepbarvar{66} + 4\irrepbarvar{12}
\end{equation}
which is anomaly free and consists of only totally antisymmetric irreps, to
avoid the occurrence of exotic fermions.  To see that this set contains
precisely three chiral families we consider the breaking of the \modelgroup
gauge symmetry to \SU{5}, which can be accomplished by many different patterns of which
we discuss two in the following paragraphs.

The totally antisymmetric irreps of \SU{12} decompose to \SU{5} as
\begin{equation}
    \setlength\arraycolsep{1pt}
    \begin{array}{l@{\;\rightarrow\;}rlcrlcrlcrlcrl}
        \irrep{12}     &   &\irrepvar{5} & &   &              & &   &                  & &   &                &+&  7&\irrepvar{1} \\
        \irrep{66}     &  7&\irrepvar{5} &+&   &\irrepvar{10} & &   &                  & &   &                &+& 21&\irrepvar{1} \\
        \irrep{220}    & 21&\irrepvar{5} &+&  7&\irrepvar{10} &+&   &\irrepbarvar{10}  & &   &                &+& 35&\irrepvar{1} \\
        \irrep{495}    & 35&\irrepvar{5} &+& 21&\irrepvar{10} &+&  7&\irrepbarvar{10}  &+&   &\irrepbarvar{5} &+& 35&\irrepvar{1} \\
        \irrep{792}    & 35&\irrepvar{5} &+& 35&\irrepvar{10} &+& 21&\irrepbarvar{10}  &+&  7&\irrepbarvar{5} &+& 22&\irrepvar{1} \\
        \irrep{924}    & 21&\irrepvar{5} &+& 35&\irrepvar{10} &+& 35&\irrepbarvar{10}  &+& 21&\irrepbarvar{5} &+& 14&\irrepvar{1} \\
        \irrepbar{792} & 7 &\irrepvar{5} &+& 21&\irrepvar{10} &+& 35&\irrepbarvar{10}  &+& 35&\irrepbarvar{5} &+& 22&\irrepvar{1} \\
        \irrepbar{495} &   &\irrepvar{5} &+&  7&\irrepvar{10} &+& 21&\irrepbarvar{10}  &+& 35&\irrepbarvar{5} &+& 35&\irrepvar{1} \\
        \irrepbar{220} &   &             & &   &\irrepvar{10} &+&  7&\irrepbarvar{10}  &+& 21&\irrepbarvar{5} &+& 35&\irrepvar{1} \\
        \irrepbar{66}  &   &             & &   &              & &   &\irrepbarvar{10}  &+&  7&\irrepbarvar{5} &+& 21&\irrepvar{1} \\
        \irrepbar{12}  &   &             & &   &              & &   &                  & &   &\irrepbarvar{5} &+&  7&\irrepvar{1}
    \end{array}
\end{equation}
For the irreps in \eqref{eq:AnomFreeSet} including their multiplicities we have
\begin{equation}\label{eq:AnomFreeSetAtSU5Level}
    3(\irrep{10}+\irrepbar{5}) + 238(\irrep{5}+\irrepbar{5}) + 211(\irrep{10}
    +\irrepbar{10}) + 487\irrepvar{1}
\end{equation}
at the \SU{5} level with three massless chiral families in
$3(\irrep{10}+\irrepbar{5})$. Vectorlike pairs of $(\irrep{5}+\irrepbar{5})$
and $(\irrep{10}+\irrepbar{10})$ as well as \SU{5} singlet fermions $\irrepvar{1}$
acquire masses at the \SU{5} unification scale. Of the sterile neutrinos in
the form of \SU{5} singlet fermions we assign three to the seesaw mechanism.
The three massless chiral families will acquire mass via the Higgs mechanism
at the electroweak scale.

We comment further here on the spontaneous symmetry breaking from \SU{12} to
\SU5 and then on to the standard model gauge group by discussing two of the
several possible patterns of symmetry breaking. Note that since our model is
supersymmetric above ${\sim}10^{11}\GeV$, one must investigate spontaneous
symmetry breaking via the superpotential. For this purpose there already exists
an analysis of the spontaneous symmetry breaking in \SU{N} models due to VEVs
for chiral superfields in the adjoint and totally antisymmetric tensor
irreps~\cite{Frampton:1981pf,Frampton:1982mj,Buccella:1982nx}. It is straightforward to show
that a single adjoint can break $\SU{N}{\to}\SU{N{-}n}{\otimes}
\SU{n}{\otimes} \U1$ and preserve supersymmetry, except when $n{=}N/2$. Hence we can
break $\SU{12}{\to}\SU{5}{\otimes} \SU{7}{\otimes} \U1$ with a single
\higgs{143}. Adding four more adjoints we can break to $\SU{5}{\otimes}
\U1^7$ and keep supersymmetry unbroken. Finally another adjoint can break \SU{5}
to $\SU{3}_\text{C}{\otimes}\SU{2}_\text{L}{\otimes}\U1_\text{Y}$. One can check
that the addition of \higgs{143} adjoint scalars does not upset the patterns
of masses and mixings we have established in our \SU{12} model. The safest way
to proceed further is to keep all the \U1's unbroken until we reach the SUSY
breaking scale where a set of singlet VEVs coming from the antisymmetric tensor
irreps with charges under the various \U1's then breaks all the \U1's except
$\U1_\text{Y}$. These low scale VEVs for components of the antisymmetric tensor
irreps will also not impact the masses and mixings. Hence we are left with
$\SU{3}_\text{C}{\otimes} \SU{2}_\text{L}{\otimes} \U1_\text{Y}$ at the SUSY
breaking scale. This procedure is rather generic and should work for most if not
all of the models in the scan.
\pagebreak

A second somewhat more appealing and economical, but less generic, approach is
to use a set of scalars coming directly from the antisymmetric chiral superfield
irreps to break \SU{12} directly to \SU{5} and then use a single adjoint to
break \SU{5} to $\SU{3}_\text{C}{\otimes}\SU{2}_\text{L}{\otimes}\U1_\text{Y}$.
This can be accomplished if the set of antisymmetric chiral superfield VEVs has
vanishing total Dynkin weight~\cite{Frampton:1981pf,Frampton:1982mj}.  At least
some of these VEVs would be expected to be in the same \SU{12} irreps as the
quark and lepton families, but this would not necessarily be so at the \SU{5}
level. Sequestering the families from the VEVs at the \SU{5} level would avoid
some technical difficulties, but the new VEVs would still leave the model in
danger of disrupted masses and mixings. This approach would necessarily require
a full analysis for each model in the scan. For the rest of this work we follow
the first more generic approach to avoid these complications.

\subsection{Fermion Assignments, Higgs and Massive Fermions}
\label{ssec:Assignments}
A successful assignment of the three \SU{5} chiral families and singlet
right-handed massive neutrinos to the \modelgroup irreps follows:
\begin{equation}\label{eq:FermionAssignments}
    \begin{array}{ll@{\,\to\:}l}
    \text{First Family}
                & (\irrep{10})\irrep{495}_1      & u_\text{L}, u^c_\text{L}, d_\text{L}, e^c_\text{L}\\
                & (\irrepbar{5})\irrepbar{66}_1  & d^c_\text{L}, e_\text{L}, \nu_{1,\text{L}}\\
                & (\irrep{1})\irrepbar{792}_1    & N^c_{1,\text{L}}\\[1mm]
    \text{Second Family}
                & (\irrep{10})\irrepbar{792}_2   & c_\text{L}, c^c_\text{L}, s_\text{L}, \mu^c_\text{L} \\
                & (\irrepbar{5})\irrepbar{792}_2 & s^c_\text{L}, \mu_\text{L}, \nu_{2,\text{L}} \\
                & (\irrep{1})\irrepbar{220}_2    & N^c_{2,\text{L}}\\[1mm]
    \text{Third Family}
                & (\irrep{10})\irrepbar{220}_3   & t_\text{L}, t^c_\text{L}, b_\text{L}, \tau^c_\text{L}  \\
                & (\irrepbar{5})\irrepbar{792}_3 & b^c_\text{L}, \tau_\text{L}, \nu_{3,\text{L}}  \\
                & (\irrep{1})\irrepbar{12}_3     & N^c_{3,\text{L}}
    \end{array}\quad\;
\end{equation}
with five unassigned \irrep{495}'s, two unassigned \irrepbar{220}'s and three
unassigned \irrepbar{12} irreps, as required by anomaly cancellation, regarded
as massive fields decoupled below the \SU{5} GUT scale as in~\eqref{eq:AnomFreeSetAtSU5Level}.

The model uses two conjugated Higgs representations containing the electroweak
VEV, \irrep{5} and \irrepbar{5} at the \SU{5} level, which contain the SM Higgs
doublet and its conjugate when \SU{5} is broken via an adjoint Higgs. Two
additional conjugate Higgs pairs containing \SU{5} singlet VEVs and two massive
fermion pairs are needed for the higher dimensional Yukawa couplings. As
explained above and discussed in detail in
Sec.~\ref{ssec:ThreeFamiliesEmbedding}, a \higgs{143} of \SU{12} and a
\higgs{24} of \SU{5} are needed for the symmetry breaking, where the latter may
be embedded in the \higgs{143}. Four more adjoints for complete \SU{7} breaking
are not displayed here. To summarize, our scalar and massive fermion content is:
\begin{equation}
    \begin{array}{l@{\,}l@{\quad\quad}l}
        \multicolumn{2}{l}{\text{Higgs bosons}}& \text{Massive fermions}\\
        (\irrep{5})\higgs{924},& (\irrepbar{5})\higgs{924}, & \irrep{220}{\times}  \irrepbar{220},\\
        (\irrep{1})\higgs{66}, & (\irrep{1})\higgsbar{66},  & \irrep{792}{\times}  \irrepbar{792}\\
        (\irrep{1})\higgs{220},& (\irrep{1})\higgsbar{220}, & \\
        (\irrep{24})\higgs{143}
    \end{array}
\end{equation}

\subsection{Yukawa Interactions}
\label{ssec:YukawaInteractions}
\begin{table*}[!t]
\renewcommand\boldirrep\relax
\setlength{\arraycolsep}{1pt}
\begin{ruledtabular}
\begin{tabularx}{\textwidth}{lX}\\[-9pt]
\twocolumnsbold{Up-Type Quark Mass-Term Diagrams}\\[3pt]
\textbf{Dim 4:} &
$\begin{array}[t]{ll}
    \UpType{3}{3} & (\irrep{10})\irrepbar{220}_3\ldot(\irrep{5})\higgs{924}\ldot(\irrep{10})\irrepbar{220}_3\\
\end{array}$\\
\textbf{Dim 5:} &
$\begin{array}[t]{ll}
    \UpType{2}{3} & (\irrep{10})\irrepbar{792}_2\ldot(\irrep{1})\higgs{66}\ldot\massivefermionpair{(\irrepbar{10})\irrep{220}}{(\irrep{10})\irrepbar{220}}\ldot(\irrep{5})\higgs{924}\ldot(\irrep{10})\irrepbar{220}_3\\
    \UpType{3}{2} & (\irrep{10})\irrepbar{220}_3\ldot(\irrep{5})\higgs{924}\ldot\massivefermionpair{(\irrep{10})\irrepbar{220}}{(\irrepbar{10})\irrep{220}}\ldot(\irrep{1})\higgs{66}\ldot(\irrep{10})\irrepbar{792}_2\\
\end{array}$\\
\textbf{Dim 6:} &
$\begin{array}[t]{ll}
    \UpType{1}{3} & (\irrep{10})\irrep{495}_1\ldot(\irrep{1})\higgs{220}\ldot\massivefermionpair{(\irrepbar{10})\irrep{792}}{(\irrep{10})\irrepbar{792}}\ldot(\irrep{1})\higgs{66}\ldot\massivefermionpair{(\irrepbar{10})\irrep{220}}{(\irrep{10})\irrepbar{220}}\ldot(\irrep{5})\higgs{924}\ldot(\irrep{10})\irrepbar{220}_3\\
    \UpType{3}{1} & (\irrep{10})\irrepbar{220}_3\ldot(\irrep{5})\higgs{924}\ldot\massivefermionpair{(\irrep{10})\irrepbar{220}}{(\irrepbar{10})\irrep{220}}\ldot(\irrep{1})\higgs{66}\ldot\massivefermionpair{(\irrep{10})\irrepbar{792}}{(\irrepbar{10})\irrep{792}}\ldot(\irrep{1})\higgs{220}\ldot(\irrep{10})\irrep{495}_1\\
    \UpType{2}{2} & (\irrep{10})\irrepbar{792}_2\ldot(\irrep{1})\higgs{66}\ldot\massivefermionpair{(\irrepbar{10})\irrep{220}}{(\irrep{10})\irrepbar{220}}\ldot(\irrep{5})\higgs{924}\ldot\massivefermionpair{(\irrep{10})\irrepbar{220}}{(\irrepbar{10})\irrep{220}}\ldot(\irrep{1})\higgs{66}\ldot(\irrep{10})\irrepbar{792}_2\\
\end{array}$\\
\textbf{Dim 7:} &
$\begin{array}[t]{ll}
    \UpType{1}{2} & (\irrep{10})\irrep{495}_1\ldot(\irrep{1})\higgs{220}\ldot\massivefermionpair{(\irrepbar{10})\irrep{792}}{(\irrep{10})\irrepbar{792}}\ldot(\irrep{1})\higgs{66}\ldot\massivefermionpair{(\irrepbar{10})\irrep{220}}{(\irrep{10})\irrepbar{220}}\ldot(\irrep{5})\higgs{924}\ldot\massivefermionpair{(\irrep{10})\irrepbar{220}}{(\irrepbar{10})\irrep{220}}\ldot(\irrep{1})\higgs{66}\ldot(\irrep{10})\irrepbar{792}_2\\
    \UpType{2}{1} & (\irrep{10})\irrepbar{792}_2\ldot(\irrep{1})\higgs{66}\ldot\massivefermionpair{(\irrepbar{10})\irrep{220}}{(\irrep{10})\irrepbar{220}}\ldot(\irrep{5})\higgs{924}\ldot\massivefermionpair{(\irrep{10})\irrepbar{220}}{(\irrepbar{10})\irrep{220}}\ldot(\irrep{1})\higgs{66}\ldot\massivefermionpair{(\irrep{10})\irrepbar{792}}{(\irrepbar{10})\irrep{792}}\ldot(\irrep{1})\higgs{220}\ldot(\irrep{10})\irrep{495}_1\\
\end{array}$\\
\textbf{Dim 8:} &
$\begin{array}[t]{ll}
    \UpType{1}{1} & (\irrep{10})\irrep{495}_1\ldot(\irrep{1})\higgs{220}\ldot\massivefermionpair{(\irrepbar{10})\irrep{792}}{(\irrep{10})\irrepbar{792}}\ldot(\irrep{1})\higgs{66}\ldot\massivefermionpair{(\irrepbar{10})\irrep{220}}{(\irrep{10})\irrepbar{220}}\ldot(\irrep{5})\higgs{924}\ldot\massivefermionpair{(\irrep{10})\irrepbar{220}}{(\irrepbar{10})\irrep{220}}\\
    &\ldot(\irrep{1})\higgs{66}\ldot\massivefermionpair{(\irrep{10})\irrepbar{792}}{(\irrepbar{10})\irrep{792}}\ldot(\irrep{1})\higgs{220}\ldot(\irrep{10})\irrep{495}_1\\
\end{array}$\\[14pt]
\twocolumnsbold{Down-Type Quark Mass-Term Diagrams}\\[3pt]
\textbf{Dim 5:} &
$\begin{array}[t]{ll}
    \DownType{3}{2} & (\irrep{10})\irrepbar{220}_3\ldot(\irrepbar{5})\higgs{924}\ldot\massivefermionpair{(\irrepbar{5})\irrepbar{220}}{(\irrep{5})\irrep{220}}\ldot(\irrep{1})\higgs{66}\ldot(\irrepbar{5})\irrepbar{792}_2\\
    \DownType{3}{3} & (\irrep{10})\irrepbar{220}_3\ldot(\irrepbar{5})\higgs{924}\ldot\massivefermionpair{(\irrepbar{5})\irrepbar{220}}{(\irrep{5})\irrep{220}}\ldot(\irrep{1})\higgs{66}\ldot(\irrepbar{5})\irrepbar{792}_3\\
\end{array}$\\
\textbf{Dim 6:} &
$\begin{array}[t]{ll}
    \DownType{3}{1} & (\irrep{10})\irrepbar{220}_3\ldot(\irrepbar{5})\higgs{924}\ldot\massivefermionpair{(\irrepbar{5})\irrepbar{220}}{(\irrep{5})\irrep{220}}\ldot(\irrep{1})\higgs{66}\ldot\massivefermionpair{(\irrepbar{5})\irrepbar{792}}{(\irrep{5})\irrep{792}}\ldot(\irrep{1})\higgsbar{220}\ldot(\irrepbar{5})\irrepbar{66}_1\\
    \DownType{2}{2} & (\irrep{10})\irrepbar{792}_2\ldot(\irrep{1})\higgs{66}\ldot\massivefermionpair{(\irrepbar{10})\irrep{220}}{(\irrep{10})\irrepbar{220}}\ldot(\irrepbar{5})\higgs{924}\ldot\massivefermionpair{(\irrepbar{5})\irrepbar{220}}{(\irrep{5})\irrep{220}}\ldot(\irrep{1})\higgs{66}\ldot(\irrepbar{5})\irrepbar{792}_2\\
    \DownType{2}{3} & (\irrep{10})\irrepbar{792}_2\ldot(\irrep{1})\higgs{66}\ldot\massivefermionpair{(\irrepbar{10})\irrep{220}}{(\irrep{10})\irrepbar{220}}\ldot(\irrepbar{5})\higgs{924}\ldot\massivefermionpair{(\irrepbar{5})\irrepbar{220}}{(\irrep{5})\irrep{220}}\ldot(\irrep{1})\higgs{66}\ldot(\irrepbar{5})\irrepbar{792}_3\\
\end{array}$\\
\textbf{Dim 7:} &
$\begin{array}[t]{ll}
    \DownType{1}{2} & (\irrep{10})\irrep{495}_1\ldot(\irrep{1})\higgs{220}\ldot\massivefermionpair{(\irrepbar{10})\irrep{792}}{(\irrep{10})\irrepbar{792}}\ldot(\irrep{1})\higgs{66}\ldot\massivefermionpair{(\irrepbar{10})\irrep{220}}{(\irrep{10})\irrepbar{220}}\ldot(\irrepbar{5})\higgs{924}\ldot\massivefermionpair{(\irrepbar{5})\irrepbar{220}}{(\irrep{5})\irrep{220}}\ldot(\irrep{1})\higgs{66}\ldot(\irrepbar{5})\irrepbar{792}_2\\
    \DownType{2}{1} & (\irrep{10})\irrepbar{792}_2\ldot(\irrep{1})\higgs{66}\ldot\massivefermionpair{(\irrepbar{10})\irrep{220}}{(\irrep{10})\irrepbar{220}}\ldot(\irrepbar{5})\higgs{924}\ldot\massivefermionpair{(\irrepbar{5})\irrepbar{220}}{(\irrep{5})\irrep{220}}\ldot(\irrep{1})\higgs{66}\ldot\massivefermionpair{(\irrepbar{5})\irrepbar{792}}{(\irrep{5})\irrep{792}}\ldot(\irrep{1})\higgsbar{220}\ldot(\irrepbar{5})\irrepbar{66}_1\\
    \DownType{1}{3} & (\irrep{10})\irrep{495}_1\ldot(\irrep{1})\higgs{220}\ldot\massivefermionpair{(\irrepbar{10})\irrep{792}}{(\irrep{10})\irrepbar{792}}\ldot(\irrep{1})\higgs{66}\ldot\massivefermionpair{(\irrepbar{10})\irrep{220}}{(\irrep{10})\irrepbar{220}}\ldot(\irrepbar{5})\higgs{924}\ldot\massivefermionpair{(\irrepbar{5})\irrepbar{220}}{(\irrep{5})\irrep{220}}\ldot(\irrep{1})\higgs{66}\ldot(\irrepbar{5})\irrepbar{792}_3\\
\end{array}$\\
\textbf{Dim 8:} &
$\begin{array}[t]{ll}
    \DownType{1}{1} & (\irrep{10})\irrep{495}_1\ldot(\irrep{1})\higgs{220}\ldot\massivefermionpair{(\irrepbar{10})\irrep{792}}{(\irrep{10})\irrepbar{792}}\ldot(\irrep{1})\higgs{66}\ldot\massivefermionpair{(\irrepbar{10})\irrep{220}}{(\irrep{10})\irrepbar{220}}\ldot(\irrepbar{5})\higgs{924}\ldot\massivefermionpair{(\irrepbar{5})\irrepbar{220}}{(\irrep{5})\irrep{220}}\\
    &\ldot(\irrep{1})\higgs{66}\ldot\massivefermionpair{(\irrepbar{5})\irrepbar{792}}{(\irrep{5})\irrep{792}}\ldot(\irrep{1})\higgsbar{220}\ldot(\irrepbar{5})\irrepbar{66}_1\\
\end{array}$\\
\end{tabularx}
\end{ruledtabular}
\caption{\label{tab:UpAndDownTypeMassTermDiagrams}Leading order up- and down-type quark diagrams for each matrix element abbreviated as discussed in Sec.~\ref{ssec:YukawaInteractions}.}
\end{table*}

By construction, the only renormalizable, dimension four Yukawa interaction is
the top-quark mass term denoted as \textbf{U33}. At the \SU{5} level the top-quark mass
term is $\fermion{10}{3}\fermion{10}{3}\higgs{5}$,arising from
$\fermionbar{220}{3}\fermionbar{220}{3}\higgs{924}$ at the \modelgroup level,
both containing singlets under their gauge groups, with the corresponding
Feynman diagram
\begin{equation}
    \textbf{U33:}\quad\raisebox{-.45\height}{\includegraphics[scale=0.90]{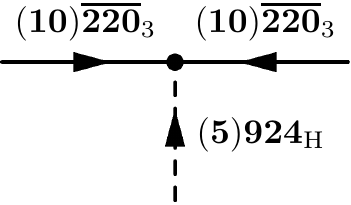}}\qquad
\end{equation}
which displays both the \SU{5}, in parentheses, followed by the \modelgroup
multiplets. After spontaneous symmetry breaking including the electroweak
symmetry, the top-quark mass term becomes: $\hu{33} v\,t^T_\text{L}t^c_\text{L}$.

All other mass terms in the full theory involve at least one mass insertion of
a heavy-fermion pair and one Higgs with an \SU{5} VEV. The corresponding
tree-level diagrams are constructed by placing the fermion multiplets at both
ends and assembling one Higgs containing the electroweak VEV and, depending
on the dimension, one or more Higgs with an \SU{5} singlet VEV and one or more
massive fermions in (massive-)fermion-massive-fermion-Higgs vertices that
individually form \modelgroup as well as \SU{5} singlets. The whole mass term
thus contains an \modelgroup and \SU{5} singlet automatically. As an
instructive example we give the bottom-quark mass term diagram (\textbf{D33}), which
will be of dimension 5 after integrating out the massive fermions:
\begin{equation}\label{eq:BottomQuarkMassTermDiagram}
    \hspace{-6pt}\textbf{D33:}\;\raisebox{-.45\height}{\includegraphics[scale=0.90]{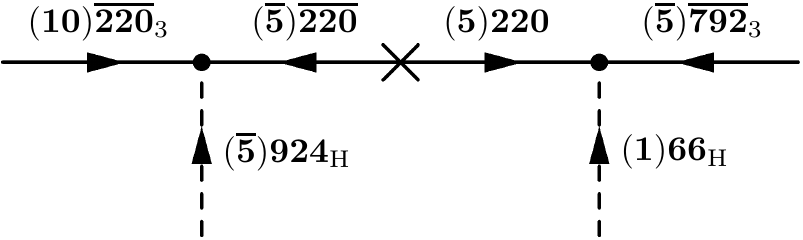}}\hspace{-9pt}
\end{equation}
We list all leading order diagrams for the quark and charged lepton matrix
elements in Table~\ref{tab:UpAndDownTypeMassTermDiagrams} using a short-hand
notation for the Feynman diagrams, which abbreviates~\eqref{eq:BottomQuarkMassTermDiagram} to
\begin{equation}
    (\irrep{10})\fermionbar{220}{3}\ldot(\irrepbar{5})\higgs{924}\ldot
  \massivefermionpair{(\irrepbar{5})\irrepbar{220}}{(\irrep{5})\irrep{220}}
  \ldot(\irrep{1})\higgs{66}\ldot(\irrepbar{5})\fermionbar{792}{3}.
\end{equation}
After integrating out massive fermions the bottom-quark mass term becomes
\begin{equation}
    (\irrep{10})\fermionbar{220}{3}
    (\irrepbar{5})\higgs{924}
    (\irrep{1})\higgs{66}
    (\irrepbar{5})\fermionbar{792}{3},
\end{equation}
and after spontaneous symmetry breaking including the electroweak one:
$\hd{33}\epsilon v\, b^T_\text{L} b^c_\text{L}$.
Note that only one diagram for each matrix element appears at leading
order, which is not self-evident in our model setup.
\pagebreak

We have defined the mass contributions in Table
\ref{tab:UpAndDownTypeMassTermDiagrams} with the left-handed fields to the left
and the left-handed conjugate fields to the right. As can be seen from
Eq.~\eqref{eq:FermionAssignments} the left- and right-handed components of the
charged leptons are flipped assignments compared to the down-type quark
components, due to the breaking of the underlying \SU{5} to the SM gauge group.
The corresponding diagrams for the charged leptons are then just the transpose
of those listed for the down quarks, since no \higgs{143} contributions appear
in the diagrams.

\subsubsection{\bf Quark Masses and Mixings}
\label{sssec:QuarkMassesAndMixings}
\begin{table*}[!t]
\renewcommand\boldirrep\relax
\setlength{\arraycolsep}{1pt}
\begin{ruledtabular}
\begin{tabularx}{\textwidth}{lX}\\[-9pt]
\twocolumnsbold{Dirac-Neutrino Mass-Term Diagrams}\\[3pt]
\textbf{Dim 4:} &
$\begin{array}[t]{ll}
    \Dirac{2}{3} & (\irrepbar{5})\irrepbar{792}_2\ldot(\irrep{5})\higgs{924}\ldot(\irrep{1})\irrepbar{12}_3\\
    \Dirac{3}{3} & (\irrepbar{5})\irrepbar{792}_3\ldot(\irrep{5})\higgs{924}\ldot(\irrep{1})\irrepbar{12}_3\\
\end{array}$\\
\textbf{Dim 5:} &
$\begin{array}[t]{ll}
    \Dirac{1}{3} & (\irrepbar{5})\irrepbar{66}_1\ldot(\irrep{1})\higgsbar{220}\ldot\massivefermionpair{(\irrep{5})\irrep{792}}{(\irrepbar{5})\irrepbar{792}}\ldot(\irrep{5})\higgs{924}\ldot(\irrep{1})\irrepbar{12}_3\\
    \Dirac{2}{2} & (\irrepbar{5})\irrepbar{792}_2\ldot(\irrep{1})\higgs{66}\ldot\massivefermionpair{(\irrep{5})\irrep{220}}{(\irrepbar{5})\irrepbar{220}}\ldot(\irrep{5})\higgs{924}\ldot(\irrep{1})\irrepbar{220}_2\\
    \Dirac{3}{2} & (\irrepbar{5})\irrepbar{792}_3\ldot(\irrep{1})\higgs{66}\ldot\massivefermionpair{(\irrep{5})\irrep{220}}{(\irrepbar{5})\irrepbar{220}}\ldot(\irrep{5})\higgs{924}\ldot(\irrep{1})\irrepbar{220}_2\\
\end{array}$\\
\textbf{Dim 6:} &
$\begin{array}[t]{ll}
    \Dirac{1}{2} & (\irrepbar{5})\irrepbar{66}_1\ldot(\irrep{1})\higgsbar{220}\ldot\massivefermionpair{(\irrep{5})\irrep{792}}{(\irrepbar{5})\irrepbar{792}}\ldot(\irrep{1})\higgs{66}\ldot\massivefermionpair{(\irrep{5})\irrep{220}}{(\irrepbar{5})\irrepbar{220}}\ldot(\irrep{5})\higgs{924}\ldot(\irrep{1})\irrepbar{220}_2\\
    \Dirac{2}{1} & (\irrepbar{5})\irrepbar{792}_2\ldot(\irrep{1})\higgs{66}\ldot\massivefermionpair{(\irrep{5})\irrep{220}}{(\irrepbar{5})\irrepbar{220}}\ldot(\irrep{5})\higgs{924}\ldot\massivefermionpair{(\irrep{1})\irrepbar{220}}{(\irrep{1})\irrep{220}}\ldot(\irrep{1})\higgs{66}\ldot(\irrep{1})\irrepbar{792}_1\\
    \Dirac{3}{1} & (\irrepbar{5})\irrepbar{792}_3\ldot(\irrep{1})\higgs{66}\ldot\massivefermionpair{(\irrep{5})\irrep{220}}{(\irrepbar{5})\irrepbar{220}}\ldot(\irrep{5})\higgs{924}\ldot\massivefermionpair{(\irrep{1})\irrepbar{220}}{(\irrep{1})\irrep{220}}\ldot(\irrep{1})\higgs{66}\ldot(\irrep{1})\irrepbar{792}_1\\
\end{array}$\\
\textbf{Dim 7:} &
$\begin{array}[t]{ll}
    \Dirac{1}{1} & (\irrepbar{5})\irrepbar{66}_1\ldot(\irrep{1})\higgsbar{220}\ldot\massivefermionpair{(\irrep{5})\irrep{792}}{(\irrepbar{5})\irrepbar{792}}\ldot(\irrep{1})\higgs{66}\ldot\massivefermionpair{(\irrep{5})\irrep{220}}{(\irrepbar{5})\irrepbar{220}}\ldot(\irrep{5})\higgs{924}\ldot\massivefermionpair{(\irrep{1})\irrepbar{220}}{(\irrep{1})\irrep{220}}\ldot(\irrep{1})\higgs{66}\ldot(\irrep{1})\irrepbar{792}_1\\
\end{array}$\\[4pt]
\twocolumnsbold{Majorana-Neutrino Mass-Term Diagrams}\\[3pt]
\textbf{Dim 4:} &
$\begin{array}[t]{ll}
    \Majorana{1}{1} & (\irrep{1})\irrepbar{792}_1\ldot(\irrep{1})\higgsbar{66}\ldot(\irrep{1})\irrepbar{792}_1\\
    \Majorana{3}{3} & (\irrep{1})\irrepbar{12}_3\ldot(\irrep{1})\higgs{66}\ldot(\irrep{1})\irrepbar{12}_3\\
\end{array}$\\
\textbf{Dim 5:} &
$\begin{array}[t]{ll}
    \Majorana{1}{2} & (\irrep{1})\irrepbar{792}_1\ldot(\irrep{1})\higgsbar{66}\ldot\massivefermionpair{(\irrep{1})\irrepbar{792}}{(\irrep{1})\irrep{792}}\ldot(\irrep{1})\higgsbar{66}\ldot(\irrep{1})\irrepbar{220}_2\\
    \Majorana{2}{1} & (\irrep{1})\irrepbar{220}_2\ldot(\irrep{1})\higgsbar{66}\ldot\massivefermionpair{(\irrep{1})\irrep{792}}{(\irrep{1})\irrepbar{792}}\ldot(\irrep{1})\higgsbar{66}\ldot(\irrep{1})\irrepbar{792}_1\\
\end{array}$\\
\textbf{Dim 6:} &
$\begin{array}[t]{ll}
    \Majorana{1}{3} & (\irrep{1})\irrepbar{792}_1\ldot(\irrep{1})\higgsbar{66}\ldot\massivefermionpair{(\irrep{1})\irrepbar{792}}{(\irrep{1})\irrep{792}}\ldot(\irrep{1})\higgsbar{66}\ldot\massivefermionpair{(\irrep{1})\irrepbar{220}}{(\irrep{1})\irrep{220}}\ldot(\irrep{1})\higgsbar{66}\ldot(\irrep{1})\irrepbar{12}_3\\
    \Majorana{3}{1} & (\irrep{1})\irrepbar{12}_3\ldot(\irrep{1})\higgsbar{66}\ldot\massivefermionpair{(\irrep{1})\irrep{220}}{(\irrep{1})\irrepbar{220}}\ldot(\irrep{1})\higgsbar{66}\ldot\massivefermionpair{(\irrep{1})\irrep{792}}{(\irrep{1})\irrepbar{792}}\ldot(\irrep{1})\higgsbar{66}\ldot(\irrep{1})\irrepbar{792}_1\\
    \Majorana{2}{2} & (\irrep{1})\irrepbar{220}_2\ldot(\irrep{1})\higgsbar{66}\ldot\massivefermionpair{(\irrep{1})\irrep{792}}{(\irrep{1})\irrepbar{792}}\ldot(\irrep{1})\higgsbar{66}\ldot\massivefermionpair{(\irrep{1})\irrepbar{792}}{(\irrep{1})\irrep{792}}\ldot(\irrep{1})\higgsbar{66}\ldot(\irrep{1})\irrepbar{220}_2\\
\end{array}$\\
\textbf{Dim 7:} &
$\begin{array}[t]{ll}
    \Majorana{2}{3} & (\irrep{1})\irrepbar{220}_2\ldot(\irrep{1})\higgsbar{66}\ldot\massivefermionpair{(\irrep{1})\irrep{792}}{(\irrep{1})\irrepbar{792}}\ldot(\irrep{1})\higgsbar{66}\ldot\massivefermionpair{(\irrep{1})\irrepbar{792}}{(\irrep{1})\irrep{792}}\ldot(\irrep{1})\higgsbar{66}\ldot\massivefermionpair{(\irrep{1})\irrepbar{220}}{(\irrep{1})\irrep{220}}\ldot(\irrep{1})\higgsbar{66}\ldot(\irrep{1})\irrepbar{12}_3\\
    \Majorana{3}{2} & (\irrep{1})\irrepbar{12}_3\ldot(\irrep{1})\higgsbar{66}\ldot\massivefermionpair{(\irrep{1})\irrep{220}}{(\irrep{1})\irrepbar{220}}\ldot(\irrep{1})\higgsbar{66}\ldot\massivefermionpair{(\irrep{1})\irrep{792}}{(\irrep{1})\irrepbar{792}}\ldot(\irrep{1})\higgsbar{66}\ldot\massivefermionpair{(\irrep{1})\irrepbar{792}}{(\irrep{1})\irrep{792}}\ldot(\irrep{1})\higgsbar{66}\ldot(\irrep{1})\irrepbar{220}_2\\
\end{array}$\\
\end{tabularx}
\end{ruledtabular}
\caption{\label{tab:DiracAndMajoranaNeutrinoMassTermDiagrams}Leading order Dirac- and Majorana-neutrino diagrams for each matrix element abbreviated as discussed in Sec.~\ref{ssec:YukawaInteractions}.}
\end{table*}
Each mass term in Table~\ref{tab:UpAndDownTypeMassTermDiagrams} is accompanied
by a coupling constant, which is assumed to be of order one at the
\modelgroup unification scale, as naturalness predicts. In Sec.~\ref{ssec:Phenomenology}
we will perform a fit to data for masses and mixings,
where these coupling constants constitute the fit parameters.
The coupling constants, also called ``prefactors'', are denoted by $\hu{ij}$
and $\hd{ij}$ for the up- and down-type quark mass terms, $\hl{ij}$ for the
charged-lepton mass terms and $\hmn{ij}$ and $\hdn{ij}$ for the Majorana- and
Dirac-neutrino mass terms, with $i,j{=} 1,2,3$.

The number of Higgs bosons with \SU{5} singlet VEVs for each mass term tells us the
exponent of the parameter $\epsilon$ occurring after \SU{5} symmetry breaking
to the SM gauge group. We can thus derive the up-type, down-type and
charged-lepton mass matrices with the coefficients of the effective mass
operators involving the prefactors $\hu{ij}$, $\hd{ij}$ and $\hl{ij}$,
respectively.

As explained above, due to the \SU5 breaking to the SM gauge
group, the charged-lepton mass matrix will be the transpose of the down-type
quark mass matrix, which also holds true for its prefactors, $\hl{ij}{=}\hd{ji}$.
This is true to the extent that no adjoint Higgs bosons with VEVs pointing
in the $B{-}L$ direction are present which would modify this transpose
structure~\cite{Babu:1995hr}.  As such, the Yukawa coupling matrices are then given by
\begin{equation}
    \begin{aligned}
        M_\text{U} =&
        \begin{pmatrix}
            \hu{11}\epsilon^4 & \hu{12}\epsilon^3  & \hu{13}\epsilon^2  \\
            \hu{12}\epsilon^3 & \hu{22}\epsilon^2  & \hu{23}\epsilon    \\
            \hu{13}\epsilon^2 & \hu{23}\epsilon    & \hu{33}            \\
        \end{pmatrix}\!v\:,\\
        M_\text{D} =&
        \begin{pmatrix}
            \hd{11}\epsilon^4 & \hd{12}\epsilon^3  & \hd{13}\epsilon^3  \\
            \hd{21}\epsilon^3 & \hd{22}\epsilon^2  & \hd{23}\epsilon^2  \\
            \hd{31}\epsilon^2 & \hd{32}\epsilon    & \hd{33}\epsilon    \\
        \end{pmatrix}\!v\:,\\
        M_\text{L} =&
        \begin{pmatrix}
            \hl{11}\epsilon^4 & \hl{12}\epsilon^3  & \hl{13}\epsilon^2  \\
            \hl{21}\epsilon^3 & \hl{22}\epsilon^2  & \hl{23}\epsilon    \\
            \hl{31}\epsilon^3 & \hl{32}\epsilon^2  & \hl{33}\epsilon    \\
        \end{pmatrix}\!v = M_\text{D}^T.
    \end{aligned}
\end{equation}

It is clear from the above that the up-quark matrix is symmetric, while the
down-quark and charged-lepton mass matrices are doubly lopsided: the terms with
$\hd{23}$ and $\hl{32}$ are suppressed by one extra power of $\epsilon$
compared with the $\hd{32}$ and $\hl{23}$ terms, respectively. For $M_\text{D}$,
for example, this implies that a larger right-handed rotation than
left-handed rotation is needed to bring the down quark matrix into diagonal
form, while the opposite is true for $M_\text{L}$~\cite{Babu:1995hr,Albright:1998vf,
Babu:2001cv}.

\subsubsection{\bf Neutrino Masses and Mixings}
\label{sssec:NeutrinoMassesAndMixings}

The assignment of heavy right-handed neutrinos to \modelgroup multiplets
containing an \SU{5} singlet allows us to explore light-neutrino masses and
mixings via the seesaw mechanism. To this end we have computed the resulting
Dirac- and the Majorana-neutrino mass terms, which are of the form
$(\hdn{ij}\epsilon^n v)\bar{\nu}_{iL}N^c_{jL}$ and
$(\hmn{ij}\epsilon^n\Lambda_\text{R}){N^c}^T_{iL}N^c_{jL}$, respectively.
The Majorana-neutrino mass terms
are constructed from only \modelgroup Higgs irreps containing \SU{5} singlet
VEVs. At the \SU{5} level, a dimension four Majorana-neutrino mass term has
the form $\irrepsub{1}{i}\irrepsub{1}{j}\higgs{1}$, while a higher dimensional
mass term involves more \SU{5} singlet Higgs. Thus the right-handed scale
$\Lambda_\text{R}$ coincides with the \SU{5} singlet VEV $\vev{1}_\SU{5}$.
The Dirac-neutrino mass term couples the left-handed neutrino in the \irrepbar{5}
at the \SU{5} level with the left-handed conjugate neutrino in the
\SU{5} singlet (see \eqref{eq:FermionAssignments}). A four-dimensional
Dirac-neutrino mass term thus has the form
$\irrepbarsub{5}{i}\irrepsub{1}{j}\higgs{5}$, while a higher dimensional
Dirac mass term involves one or more \SU{5} Higgs singlets.
The Dirac- and Majorana-neutrino mass diagrams arising from the given fermion
assignments and set of Higgs bosons and massive fermions are listed in Table
\ref{tab:DiracAndMajoranaNeutrinoMassTermDiagrams}. As for the quark and charged
lepton mass matrices, only one diagram for each matrix element appears at leading
order.

The corresponding mass matrices are:
\begin{equation}
    \begin{aligned}
        M_\text{DN} =&
        \begin{pmatrix}
            \hdn{11}\epsilon^3 & \hdn{12}\epsilon^2 & \hdn{13}\epsilon \\
            \hdn{21}\epsilon^2 & \hdn{22}\epsilon   & \hdn{23} \\
            \hdn{31}\epsilon^2 & \hdn{32}\epsilon   & \hdn{33}
        \end{pmatrix}\!v\:,\\
        M_\text{MN} =&
        \begin{pmatrix}
            \hmn{11}           & \hmn{12}\epsilon   & \hmn{13}\epsilon^2 \\
            \hmn{12}\epsilon   & \hmn{22}\epsilon^2 & \hmn{23}\epsilon^3 \\
            \hmn{13}\epsilon^2 & \hmn{23}\epsilon^3 & \hmn{33}
        \end{pmatrix}\!\Lambda_\text{R}.\\
    \end{aligned}
     \vspace*{2pt}
\end{equation}
Observe that not only are $M_\text{D}$ and $M_\text{L}$ doubly lopsided, but
$M_\text{DN}$ is as well. The symmetric light-neutrino mass matrix is obtained
via the Type I Seesaw mechanism:
\begin{equation}
    M_\nu = -M_\text{DN}M_\text{MN}^{-1}M_\text{DN}^T.
\end{equation}
In accordance with the construction of the up- and down-type quark mass
matrices, we use only the leading term in $\epsilon$ for each matrix element
of the light-neutrino mass matrix, yielding
\begin{widetext}
     \vspace*{-9pt}
    \begin{equation}
        M_\nu \approx \frac{v^2}{\Lambda_\text{R}}\times\!\!\\
        \begin{pmatrix}
            \epsilon ^2 \left(\dfrac{\hdn{12}^2 \hmn{11}}{\hmn{12}^2{-}\hmn{11} \hmn{22}}{-}\dfrac{\hdn{13}^2}{\hmn{33}}\right)
            & \epsilon  \left(\dfrac{\hdn{12} \hdn{22} \hmn{11}}{\hmn{12}^2{-}\hmn{11} \hmn{22}}{-}\dfrac{\hdn{13} \hdn{23}}{\hmn{33}}\right)
            & \epsilon  \left(\dfrac{\hdn{12} \hdn{32} \hmn{11}}{\hmn{12}^2{-}\hmn{11} \hmn{22}}{-}\dfrac{\hdn{13} \hdn{33}}{\hmn{33}}\right)\\[3mm]
            \epsilon  \left(\dfrac{\hdn{12} \hdn{22} \hmn{11}}{\hmn{12}^2{-}\hmn{11} \hmn{22}}{-}\dfrac{\hdn{13} \hdn{23}}{\hmn{33}}\right)
            & \dfrac{\hdn{22}^2 \hmn{11}}{\hmn{12}^2{-}\hmn{11} \hmn{22}}{-}\dfrac{\hdn{23}^2}{\hmn{33}}
            & \dfrac{\hdn{22} \hdn{32} \hmn{11}}{\hmn{12}^2{-}\hmn{11} \hmn{22}}{-}\dfrac{\hdn{23} \hdn{33}}{\hmn{33}} \\[3mm]
            \epsilon  \left(\dfrac{\hdn{12} \hdn{32} \hmn{11}}{\hmn{12}^2{-}\hmn{11} \hmn{22}}{-}\dfrac{\hdn{13} \hdn{33}}{\hmn{33}}\right)
            & \dfrac{\hdn{22} \hdn{32} \hmn{11}}{\hmn{12}^2{-}\hmn{11} \hmn{22}}{-}\dfrac{\hdn{23} \hdn{33}}{\hmn{33}}
            & \dfrac{\hdn{32}^2 \hmn{11}}{\hmn{12}^2{-}\hmn{11} \hmn{22}}{-}\dfrac{\hdn{33}^2}{\hmn{33}}
        \end{pmatrix}
    \end{equation}
\end{widetext}
which does not involve the prefactors $\hdn{11}$, $\hdn{21}$, $\hdn{31}$,
$\hmn{13}$ and $\hmn{23}$. These prefactors remain undetermined
by the fit described in Sec.~\ref{ssec:Phenomenology}, reducing
the number of fit parameters as opposed to using the full expression, and
thereby improving the fit convergence somewhat.

The light-neutrino mass matrix exhibits a much milder hierarchy compared to
the up-type and down-type mass matrices, as can be seen from the pattern of
powers of $\epsilon$. A mild or flat hierarchy of $M_\nu$ is conducive to
obtaining large mixing angles and similar light neutrino masses. Furthermore,
one observes that the light neutrino mass matrix obtained via the
seesaw mechanism involves the doubly lopsided Dirac neutrino mass matrix twice. The
lopsided feature of $M_\text{DN}$ is such as to require a large left-handed
rotation to bring $M_\nu$ into diagonal form.

\subsection{Phenomenology}
\label{ssec:Phenomenology}

The phenomenological implications of the model presented here are encoded in
the mass matrices. Normally the up-type, down-type, charged-lepton and
light-neutrino masses are the eigenvalues of the corresponding mass matrices
$M_\text{U}$, $M_\text{D}$, $M_\text{L}$ and $M_\nu$, but since not all of
these matrices are hermitian we diagonalize $M M^\dagger$ instead. Thus, with
left-handed rotations we obtain real and positive eigenvalues as squares of
the corresponding masses, according to
\begin{equation}
    \begin{aligned}
        \diag(m_u^2, m_c^2, m_t^2)          &= U_\text{U}^\dagger M_\text{U}M_\text{U}^\dagger U_\text{U},\\
        \diag(m_d^2, m_s^2, m_b^2)          &= U_\text{D}^\dagger M_\text{D}M_\text{D}^\dagger U_\text{D},\\
        \diag(m_e^2, m_\mu^2, m_\tau^2)     &= U_\text{L}^\dagger M_\text{L}M_\text{L}^\dagger U_\text{L},\\
        \diag(m_{\nu_1}^2, m_{\nu_2}^2, m_{\nu_3}^2) &= U_\nu^\dagger M_\nu M_\nu^\dagger U_\nu.
    \end{aligned}
\end{equation}

The Cabibbo-Kobayashi-Maskawa (CKM) matrix $V_\text{CKM}$ is calculated from
the unitary transformations $U_\text{U}$ and $U_\text{D}$ that diagonalize the
up-type and down-type mass matrices respectively:
\begin{equation}
    V_\text{CKM} = U_\text{U}^\dagger U_\text{D},
\end{equation}
encoding the mismatch of the flavor and mass eigenbases of the up-type and
down-type quarks.
The Pontecorvo-Maki-Nakagawa-Sakata (PMNS) matrix $V_\text{PMNS}$ is obtained
analogously from $U_\text{L}$ and $U_\nu$ that diagonalize the charged-lepton
mass matrix $M_\text{L}$ and the light-neutrino mass matrix $M_\nu$:
\begin{equation}
    V_\text{PMNS} = U_\text{L}^\dagger U_\nu.
\end{equation}
From the doubly lopsided natures of the three matrices, $M_\text{D},\ M_\text{L},$ and $M_\text{DN}$,
discussed earlier, we anticipate that for the mixing of the left-handed
fields described by $V_\text{CKM}$ and $V_\text{PMNS}$, small mixing angles
will appear in the former and large mixing angles in the latter.

\subsubsection{\bf Fit Setup}
\begin{table*}[t]
    \begin{tabular}{lllll}
        \textbf{Up-type masses} & \textbf{Down-Type masses} & \textbf{CKM Matrix}  \\
        $\begin{array}{@{}l@{\,=\,}l}
            m_\text{u} & 2.2\MeV \\
            m_\text{c} & 600\MeV \\
            m_\text{t} & 166\GeV
        \end{array}$ &
        $\begin{array}{@{}l@{\,=\,}l}
            m_\text{d} & 3.8\MeV  \\
            m_\text{s} & 75\MeV   \\
            m_\text{b} & 2.78\GeV
        \end{array}$  &
        $\begin{pmatrix}
             0.974 &  0.225 & 0.003 \\
            -0.225 &  0.973 & 0.041 \\
             0.009 & -0.040 & 0.999
        \end{pmatrix}$\\
        \\[-2ex]
        \textbf{Ch. Lepton masses} & \textbf{Neutrino Mass Diff.}& \textbf{PMNS Matrix}& \textbf{Mixing Angles} & \textbf{Phase}\\
        $\begin{array}{@{}l@{\,=\,}l}
            m_\text{e} & 0.501\MeV \\
            m_\mu      & 104\MeV   \\
            m_\tau     & 1.75\GeV
        \end{array}$ &
        $\begin{array}{@{}l@{\,=\,}l}
            \abs{\Delta_{21}} & 7.6{\times}10^{-5}\eVsq \\
            \abs{\Delta_{31}} & 2.4{\times}10^{-3}\eVsq\\
            \abs{\Delta_{32}} & 2.4{\times}10^{-3}\eVsq
        \end{array}$&
        $\begin{pmatrix}
             0.824 & 0.547 & -0.145 \\
            -0.500 & 0.582 & -0.641 \\
            -0.267 & 0.601 &  0.754
        \end{pmatrix}$&
        $\begin{array}{@{}l@{\,=\,}l}
            \sin^2\theta_{12} & 0.306 \\
            \sin^2\theta_{23} & 0.420\\
            \sin^2\theta_{13} & 0.021
        \end{array}$&
        $\delta\,=\,\pi$
    \end{tabular}
    \vspace{2pt}
    \caption{\label{tab:Inputs} Phenomenological data entering the fit with masses at the top-quark scale.}
    \vspace{2pt}
\end{table*}
\begin{table*}[t]
    \begin{tabular}{lllll}
        \textbf{Up-type masses} & \textbf{Down-Type masses} & \textbf{CKM Matrix}\\
        $\begin{array}{@{}l@{\,=\,}l}
            m_\text{u} & 2.1\MeV \\
            m_\text{c} & 600\MeV  \\
            m_\text{t} & 166\GeV
        \end{array}$ &
        $\begin{array}{@{}l@{\,=\,}l}
            m_\text{d} & 2.7\MeV \\
            m_\text{s}  & 90.7\MeV \\
            m_\text{b} & 2.32\GeV
        \end{array}$&
        $\begin{pmatrix}
             0.974 &  0.227 & 0.003 \\
            -0.227 &  0.973 & 0.042 \\
             0.007 & -0.042 & 0.999
        \end{pmatrix}$ \\
        \\[-2ex]
        \textbf{Ch. Lepton masses} & \textbf{Neutrino Mass Diff.} & \textbf{PMNS Matrix} & \textbf{Mixing Angles} & \textbf{Phase}\\
        $\begin{array}{@{}l@{\,=\,}l}
            m_\text{e} & 2.7\MeV \\
            m_\mu      & 90.7\MeV \\
            m_\tau     & 2.32\GeV
        \end{array}$&
        $\begin{array}{@{}l@{\,=\,}l}
            \abs{\Delta_{21}} & 7.5{\times}10^{-5}\eVsq \\
            \abs{\Delta_{31}} & 2.5{\times}10^{-3}\eVsq\\
            \abs{\Delta_{32}} & 2.4{\times}10^{-3}\eVsq
        \end{array}$&
        $\begin{pmatrix}
             0.824 & 0.548 & -0.145 \\
            -0.500 & 0.582 & -0.641 \\
            -0.267 & 0.601 &  0.754
        \end{pmatrix}$&
        $\begin{array}{@{}l@{\,=\,}l}
            \sin^2\theta_{12} & 0.306 \\
            \sin^2\theta_{23} & 0.420\\
            \sin^2\theta_{13} & 0.021
        \end{array}$ &
        $\delta\,=\,\pi$ \\
        \\[-1.5ex]
        \textbf{Heavy Neutrinos} & \textbf{Light Neutrinos} \\
        $\begin{array}{@{}l@{\,=\,}l}
            M_1 & 1.67{\times}10^{12}\GeV \\
            M_2 & 6.85{\times}10^{13}\GeV\\
            M_3 & 5.30{\times}10^{14}\GeV
        \end{array}$ &
        $\begin{array}{@{}l@{\,=\,}l}
            m_1 & 0.0\meV \\
            m_2 & 8.65\meV \\
            m_3 & 49.7\meV
        \end{array}$
    \end{tabular}
    \vspace{2pt}
    \caption{\label{tab:FitResults}Theoretical mass and mixing results obtained from the fitting procedure.}
    \vspace{2pt}
\end{table*}
The results still depend on the prefactors $\hu{ij}$, $\hd{ij}{=}\hl{ji}$,
$\hdn{ij}$ and $\hmn{ij}$. All four independent sets of prefactors are of
$\order{1}$ at the \modelgroup unification scale in our Froggatt-Nielsen
scenario. To test the model, the prefactors should be fit with data for the
masses and mixing matrices. The best fit should give reasonable theoretical
predictions, and a $\chi^2$ value serves as goodness-of-fit measure. Obviously
the data and prediction should be fit at a common scale, e.g.,\ the top-quark
mass scale. Hence, the running of the prefactors has to be calculated, and their
values from the fit run to the \modelgroup unification scale should turn out to
be of $\order{1}$.

For the fit we consider only real prefactors of the CKM and PMNS matrix
elements, to avoid too many fit parameters for a good convergence of the fit. We
have adhered to the Particle Data Group (PDG) sign convention for the CKM
matrix~\cite{Nakamura:2010zzi} but used the tri-bimaximal mixing sign convention
for the PMNS lepton mixing matrix~\cite{Harrison:2002er}. As common scale for
the fit, we choose for the top-quark scale $m_t(m_t){\simeq}166\GeV$ and use
extrapolated masses for the quarks and charged lepton masses
from~\cite{Babu:1999me}, where they have been calculated using three-loop QCD
and one-loop QED beta functions.

We use the measured values of the CKM matrix elements, with PDG sign convention,
without extrapolating to the top-quark mass scale. The renormalization group
flow of the CKM matrix is governed by the Yukawa couplings, which are small
except for the top-quark. Thus the effect is negligible, especially for the
matrix elements of the first two families, and small for the third family in the
Standard Model \cite{Balzereit:1998id}, which also holds true  for the low scale
of the \modelgroup model presented here. As data for the fit of the neutrino
sector, we use the mass squared differences of the light neutrinos and the
neutrino mixing angles obtained by a global analysis of oscillation
data~\cite{Fogli:2011qn}. The PMNS matrix entering the fit as data is computed from the
neutrino mixing angles using the PDG parametrization of the PMNS
matrix~\cite{Nakamura:2010zzi} but with the tri-bimaximal mixing sign
convention~\cite{Harrison:2002er}. Note that the 13 element of the PMNS matrix
is non-zero, as opposed to that for tri-bimaximal mixing, but in accord with the
evidence for a non-zero $\theta_{13}$~\cite{Fogli:2011qn}. A negative sign for the
13 element gives us better fit convergence and theoretical predictions than a
positive one, which coincides with the preference for the CP phase of
$\cos\delta{=}{-}1$ in~\cite{Fogli:2011qn}.

With respect to the \modelgroup unification scale of $\mathcal{O}(10^{16})\GeV$,
the scale of neutrino measurements is near the top-quark scale (${\sim}1\MeV$
for reactor and solar neutrinos and ${\sim}1\GeV$ for accelerator and
atmospheric neutrinos). We assume here that the running between the two neutrino
scales is small compared to the uncertainties in neutrino measurements.
\pagebreak

The quark and charged-lepton masses and light-neutrino mass differences, as well
as the CKM and PMNS matrix elements we use as data in the fit are listed in
Table~\ref{tab:Inputs}. The fit uses 6 quark masses, 3 charged-lepton masses, 3
light-neutrino mass squared differences, and 9 CKM and 9 PMNS matrix elements as
observations, for a total of $n_\text{data}{=}30$.

The fit parameters are the prefactors of the four mass matrices and the
right-handed scale $\Lambda_\text{R}$, i.e. $n_\text{params}{=}n_\text{prefactors}{+}1$.
Since the up-type mass matrix as well as the Majorana-neutrino mass matrix are
symmetric, they involve only 6 independent fit parameters each, while the
down-type mass matrix and the Dirac-neutrino mass matrix each contribute 9
parameters. As explained in Sec.~\ref{sssec:NeutrinoMassesAndMixings}, only the
leading order in $\epsilon$ of the light-neutrino mass matrix is used in the
fit, which does not involve 3 prefactors of the Dirac-neutrino and 2 of the
Majorana-neutrino mass matrix; thus 5 neutrino related prefactors remain
undetermined, yielding a total of $n_\text{prefactors}{=}25$ prefactors used in
the fit.

It is clear that the ratio of the \SU{5} singlet VEV to the \modelgroup
unification scale used as the basic parameter,
$\epsilon{=}\vev{1}_\SU{5}/M_{\modelgroup}{\sim}1/50$, in our model should be
determined by the fit as well. However, we observe a bad convergence of the
fit, when we allow it to vary. Thus, we were forced to fix its value
and found $\epsilon{=}1/6.5^2{=}0.0237$ to be an appropriate value in accord
with \cite{Babu:1999me}. The resulting number of degrees of freedom is then
$n_\text{dof}{=}n_\text{data}{-}n_\text{prefactors}{-}1{=}4$.

\subsubsection{\bf Fit Results}

The mass matrices with the results for the prefactors inserted are listed
below:
\begin{equation}
    \begin{aligned}
        M_\text{U} =&
        \begin{pmatrix}
            {-}1.1\epsilon^4 &    7.1\epsilon^3  & 5.6\epsilon^2  \\
               7.1\epsilon^3 & {-}6.2\epsilon^2 & {-}0.10\epsilon \\
               5.6\epsilon^2 & {-}0.10\epsilon    & {-}0.95       \\
        \end{pmatrix}\!v,\\
        M_\text{D} =&
        \begin{pmatrix}
            {-}6.3\epsilon^4  &   8.0\epsilon^3  & {-}1.9\epsilon^3 \\
            {-}4.5\epsilon^3  &   0.38\epsilon^2 & {-}1.3\epsilon^2 \\
               0.88\epsilon^2 &  {-}0.23\epsilon & {-}0.51\epsilon  \\
        \end{pmatrix}\!v,\\
        M_\text{DN} =&
        \begin{pmatrix}
            \hdn{11}\epsilon^3 & 0.21\epsilon^2  & {-}2.7\epsilon \\
            \hdn{21}\epsilon^2 & {-}0.28\epsilon & {-}0.15        \\
            \hdn{31}\epsilon^2 & 2.1\epsilon     & 0.086          \\
        \end{pmatrix}\!v,\\
        M_\text{MN} =&
        \begin{pmatrix}
            {-}0.72            & {-}1.5\epsilon     & \hmn{13}\epsilon^2 \\
            {-}1.5\epsilon     & 0.95\epsilon^2     & \hmn{23}\epsilon^3 \\
            \hmn{13}\epsilon^2 & \hmn{23}\epsilon^3 & 0.093              \\
        \end{pmatrix}\!\Lambda_\text{R},\\
        M_\nu =&
        \begin{pmatrix}
            {-}81.\epsilon^2 & {-}4.3\epsilon  &    2.4\epsilon \\
            {-}4.3\epsilon   & {-}0.25         &    0.28        \\
               2.4\epsilon   &    0.28         & {-}1.1         \\
        \end{pmatrix}\!\frac{v^2}{\Lambda_\text{R}},
    \end{aligned}
\end{equation}
with the right-handed scale determined to be $\Lambda_\text{R}{=}7.4{\times}10^{14}\GeV$
and $\Delta_{32}$ fit with $m_3{\sim}50$ meV.
As explained in Sec.~\ref{sssec:NeutrinoMassesAndMixings}, $\Lambda_\text{R}$
coincides with the \SU{5} singlet VEV, $\vev{1}_\SU{5}$, which allows us to
determine the \modelgroup unification scale from the fit to be
$M_{\modelgroup}{=}\Lambda_\text{R}/\epsilon{=}3.1{\times}10^{16}\GeV$.

The corresponding theoretical predictions for the masses and mixings are listed
in Table \ref{tab:FitResults}. The predictions are nearly in perfect agreement
with the phenomenological data entering the fit, which is due to the fact that
almost as many fit parameters as data points are used. This is reflected in the
abnormally small $\chi^2/n_\text{dof}$ and large P-value: With $\chi^2{=}0.239$
and $n_\text{dof}{=}4$ we obtain $\chi^2/n_\text{dof}{=}0.060$ and a P-value of
$\text{prob}(0.239, 4){=}0.993$. Only the lepton and down-quark masses deviate
significantly from their measured value, since our \modelgroup model forces them
to be equal which keeps the $\chi^2$ from dropping even further.

It is evident conclusions that can be drawn from the fit results for our
\modelgroup model are somewhat limited, for only a slightly different
phenomenology could be accommodated by an according shift of the prefactors. The
prefactors obtained from the fit can be considered to be of $\order{1}$ as
required by naturalness, aside from the 11 element of $M_\nu$. However,
according to the phenomenological input discussed above, their values apply
at the top-quark scale. For a complete analysis one has to run them to the
\modelgroup unification scale, where their compliance with the naturalness
paradigm is supposed to be probed. This calculation as well as a fully fledged
fit, including uncertainties, correlations, complex prefactors and the running
of the CKM matrix and of the neutrino data to the top-quark scale goes
beyond the scope of this paper.

Given the above caveats, we note that a normal mass hierarchy for the light
neutrinos is obtained with one massless neutrino.  Allowing for the sizable
reactor neutrino angle confirmed by the fit and the fully allowed ranges of the
Dirac and Majorana phases not present in our analysis, the effective mass
prediction for neutrino-less double beta decay lies in the range 1.5 - 3.7 meV.

\section{Discussion}
\label{sec:Discussion}

Most flavor symmetry models studied to date involve discrete flavor groups. A
typical model in this class based on the standard model gauge group or on a more
general \SU{N} family gauge group has an additional discrete flavor symmetry $G$
with the matter spectrum living in irreps of $G$. However, such models have several
disadvantages compared to models that have no additional discrete symmetry.

First disadvantage to having a discrete symmetry is that if it is a global
symmetry, it will be broken by gravity~
\cite{Holman:1992us,Kamionkowski:1992mf,Barr:1992qq}, and the breaking will not
in general be in the pattern one wishes to arrange for the family symmetry.

Second, it is difficult to explain the origin of a discrete symmetry in a more
fundamental theory. It could arise from breaking a gauge symmetry and avoid the
problems with gravity, but this is difficult to arrange. In that case it would
be necessary for $G$ to be anomaly-free~\cite{Luhn:2008sa,Luhn:2008sa}. Another
disadvantage of including a discrete symmetry is that when it breaks, cosmic
domain walls are produced. The walls need to be removed, and they can be
inflated away in some models, but not all. In particular, if there is a discrete
symmetry breaking after inflation, then the cosmology of the model will be
untenable.

If we go to larger $N$ to avoid $G$ as in the present work, then there is no
domain wall problem. There is usually still a magnetic monopole problem that
needs to be solved by inflation. However, this can be done at the GUT scale, and it
does not re-emerge at a lower scale. (The $SM{\otimes} G$ and $\SU{N}{\otimes} G$
models also have similar magnetic monopole problems.) So we conclude that the
cosmology of the discrete symmetry free models is typically more attractive.
Their one disadvantage is that the initial gauge group is usually larger, but
not below the GUT scale.

We see a balance between the two types of models. Including a discrete symmetry
to arrange a desired behavior for masses and mixings in $SM{\otimes} G$ and
$\SU{N}{\otimes} G$ models can be offset by increasing $N$ in pure gauge \SU{N}
models to avoid the inclusion of $G$. Since no domain wall problem or
problem with gravity arises if $G$ can be avoided, we conclude that pure gauge
family symmetric models like the \SU{12} model presented here, have several
advantages over flavor-symmetric models
that contain discrete symmetries.

In our studies of models of different \SU{N}'s we find that with increasing $N$
it is possible to obtain models with more and more desired features implemented.
Those features show not only compliance with phenomenology but additional esthetic
properties such as simplicity. Nevertheless, selecting a specific assignment of
fermions and Higgs scalars out of millions of possible assignments, because of its
ability to reproduce phenomenology, is yet another application of the anthropic
principle and is reminiscent of the string theory landscape.

\section{Summary and Conclusion}
\label{sec:SummaryAndConclusion}

We have developed a systematic computer scan for \SU{N} family and flavor
unification models that reproduce the observed fermion mass and mixing hierarchy
with higher-dimensional effective Yukawa couplings involving an extended Higgs sector.
These models are of the supersymmetric type since the higher-dimensional Yukawa
couplings stem from Froggatt-Nielson-type diagrams involving massive fermion insertions.
The three families of fermions, the massive fermions and the Higgs scalars are assigned to various
\SU{N} representations and may also involve the assignment of discrete symmetry charges.
A basic parameter in this setup is the ratio of the scale of imposed \SU{5} singlet
VEVs to the \SU{N} unification scale, denoted as $\epsilon$, with a value of roughly the
ratio of the bottom-quark to the top-quark mass, i.e.~${\sim}1/50$.

In this paper we have presented an example of an \SU{12} model obtained by our
computer scan, which does not involve any discrete flavor symmetry.  This
particular model belongs to a subset of economic \SU{12} models having only two
pairs of Higgs bosons with \SU{5} singlet VEVs besides the conjugate Higgs field
with the pair of electroweak VEVs, and two massive fermion pairs at the \SU{12}
level. However, we need several additional \SU{12} irreps for anomaly
cancellation, and owing to the large \SU{12} gauge group we predict
a host of fermions which become massive after symmetry breakdown to
\SU{5}.  Only three \SU{5} sets of $\irrep{10} + \irrepbar{5}$ fermions
remain massless down to the electroweak scale, while three \SU{5}
left-handed neutrino conjugate singlets take part in the seesaw mechanism.

The model presented has only one diagram at leading order in $\epsilon$ for each
matrix element of the five up, down, charged lepton, Dirac- and Majorana-neutrino
mass matrices. The down-type, lepton and Dirac-neutrino mass matrices are found to be
doubly lopsided.
The mass matrices involve undetermined Yukawa couplings, called ``prefactors,''
which are supposed to be of $\order{1}$ at the \SU{12} unification scale.  Being able
to compute all quark and lepton masses and mixings from their dependence on these
prefactors, we performed a simple fit to experimental data to test their naturalness
and the compliance of the model with phenomenology. We have presented here a fit
result with prefactors that can be considered of $\order{1}$ and a near to perfect
agreement of theoretical prediction with phenomenological data. In addition  our
analysis of the neutrino sector involving the type-I seesaw mechanism allows us to
determine the light-neutrino masses and thus their hierarchy, as well as the
heavy-neutrino masses and the full PMNS matrix. We find a normal hierarchy with
one light-neutrino mass being zero.

Still the predictive power of our simple analysis is limited: We used the top-quark
scale as common scale for the fit. Thus the determined values of the prefactors
apply at this scale and should be run to the \SU{12} unification scale to test
their naturalness in a rigorous analysis. We also have not included any CP
phases in the mass matrices. Furthermore, the nearly perfect agreement of
theoretical prediction with phenomenological data is due to a large number of
fit parameters, which are mostly prefactors. Besides being of $\order{1}$, there
is no a-priory estimate of their value as initial value. Since differences in
numerators and denominators of mixing-matrix elements are involved, the
uniqueness of the $\chi^2$ minimum must be doubted. In a fully fledged analysis
pull distributions generated by toy Monte Carlos clarify this aspect of the fit
quality. A rigorous analysis would also include uncertainties of experimental
data as well as estimations of the theoretical uncertainties. Nevertheless, we
believe the alternative approach to unification of families and flavors explored
here warrants further study despite the limitations of our analysis cited above.

\begin{acknowledgments}
We thank Qaisar Shafi and William Bardeen for useful discussions. The work of
RPF was supported by a fellowship within the Postdoc-Programme of the German
Academic Exchange Service (DAAD). The work of RPF and TWK was supported by US
DOE grant E-FG05-85ER40226. One of us (CHA) thanks the Fermilab Theoretical
Physics Department for its kind hospitality, where part of this work was carried
out. Fermilab is operated by Fermi Research Alliance, LLC under Contract No.
De-AC02-07CH11359 with the U.S. Department of Energy.
\end{acknowledgments}

\bibliography{references}

\end{document}